\documentstyle[prl,aps,epsf,multicol,tabularx]{revtex}
\tighten
\begin{document}
\draft
\title{CHARGE RELAXATION AND DEPHASING IN COULOMB COUPLED CONDUCTORS}
\author{Markus B\"uttiker and Andrew M. Martin}
\address{D\'epartement de Physique Th\'eorique, Universit\'e de Gen\`eve,
CH-1211 Gen\`eve 4, Switzerland}
\date{\today}
\maketitle
\begin{abstract}
The dephasing time in coupled mesoscopic conductors is caused
by the fluctuations of the dipolar charge permitted by the long
range Coulomb interaction. We relate the phase breaking time
to elementary transport coefficients which describe the dynamics
of this dipole: the capacitance, an equilibrium
charge relaxation resistance and in the presence of transport through one
of the conductors a non-equilibrium charge relaxation resistance. The
discussion is illustrated for a quantum point contact in a high magnetic
field in proximity to a quantum dot.
\end{abstract}
\begin{multicols}{2}
\narrowtext
Mesoscopic systems coupled only via the long range Coulomb
forces are of importance since one of the systems
can be used to perform measurements on the other \cite{field}.
Despite the absence of carrier transfer between the two conductors
their proximity affects the dephasing rate. Of particular interest
are {\it which path} detectors which can provide information
on the paths of a carrier in an interference
experiment \cite{buks,aleiner1,levinson}.
It is understood that
at very low temperatures the basic processes which limit the time
$\tau_{\phi}$ over which a carrier preserves its quantum mechanical
phase are electron-electron interaction processes \cite{aak,blanter}.
For the zero-dimensional conductors of interest here,
the basic process is a charge accumulation in one of the conductors
accompanied by a charge depletion in the other conductor.
The Coulomb coupling of two conductors
manifests itself in the formation of a
charge dipole and the fluctuations of this dipole
governs the dephasing process. The dynamics of this dipole,
and thus the dephasing rate,
can be characterized by elementary transport coefficients:
In the absence of an external bias
excess charge relaxes toward
its equilibrium value with an $RC$-time. In mesoscopic
conductors\cite{btp} the $RC$-time is determined by an electrochemical
capacitance $C_{\mu}$ and a charge relaxation resistance $R_q$.
In the presence of transport through one of the conductors, the charge pile-up
associated with shot noise\cite{bu90}, leads to a non-equilibrium charge
relaxation resistance\cite{plb} $R_v$. Below we relate $R_q$ and $R_v$
to the dephasing rate.

Renewed interest in dephasing 
was also generated by experiments on metallic diffusive
conductors and a suggested role of zero-point fluctuations\cite{mohant}.
We refer to the resulting discussion only with a recent item\cite{zawad}.
More closely related to our work are experiments
by Huibers et al. \cite{huibers} in which the dephasing rate
in chaotic cavities is measured. At low frequencies such cavities can be
treated as zero dimensional systems\cite{brouwer}.

Consider two mesoscopic conductors
coupled by long range Coulomb interactions.
An example of such a sys-
\begin{figure}
\narrowtext
\vspace*{0.5cm}
\epsfxsize=7cm
\centerline{\epsffile{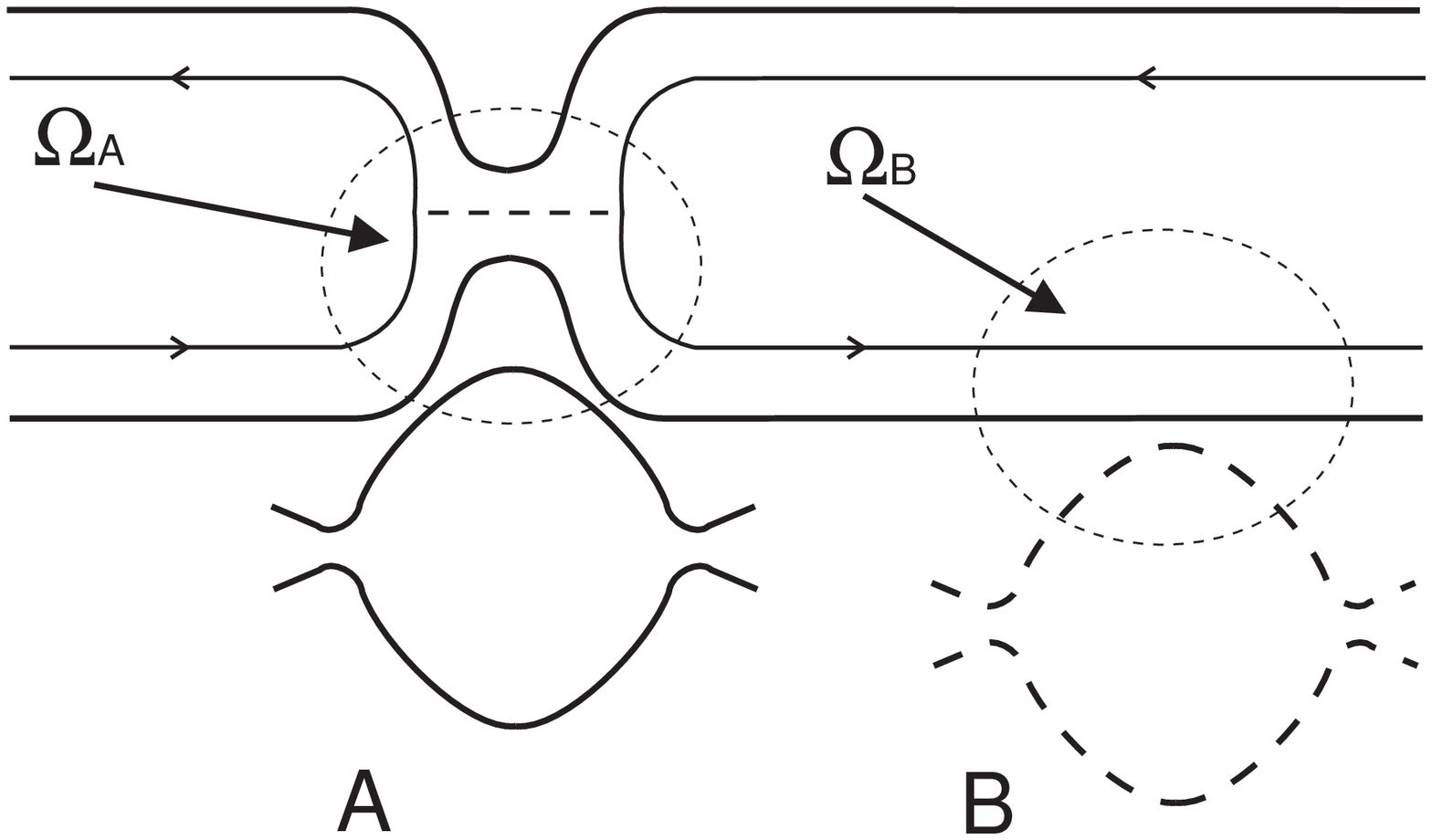}}
\vspace*{0.5cm}
\caption{ \label{qpc_geometry}
Quantum point contact coupled to a quantum dot either in position A or B.
}
\end{figure}
\noindent tem, suggested in Ref. \cite{note1}, is shown in Fig. 1.
In case A,
a quantum point contact (QPC) in a high magnetic field
is close to a quantum dot
and in case B the QPC is some distance away from a quantum dot.
First we focus on case A.
To describe the charge dynamics of such a system
we use two basic elements. First we characterize the long
range Coulomb interaction with the help of a geometrical capacitance,
much as in the literature on the Coulomb blockade.
Second the electron dynamics in each conductor $(i)$ is
described with the help of its scattering matrix,
$s^{(i)}_{\alpha\beta}(E,U_{i})$ which relates the amplitudes of incoming
currents at
contact $\beta$ to the amplitudes of the outgoing currents at
$\alpha$. The scattering matrix is a function of the energy of the carriers
and is a function of the electrostatic potential $U_{i}$ inside conductor
$i$. In case A, the total
excess charge on the conductor is of importance. In this case
the charge dynamics of the mesoscopic conductor
can be described with the help of a density of states matrix
\begin{equation}
{\cal N}^{(i)}_{\delta\gamma} = \frac{1}{2\pi i} \sum_\alpha
s^{(i) \dagger}_{\alpha\delta}\frac{ds^{(i)}_{\alpha\gamma}}{dE}.
\label{d0}
\end{equation}
Eq. (\ref{d0}) is valid in the WKB limit in which derivatives
with regard to the potential can be replaced by an energy derivative.
Eq. (\ref{d0}) are elements of the Wigner-Smith delay-time
matrix \cite{smith,note2}.
Later, we consider also situations in which energy derivatives
are not sufficient.
The diagonal elements of this matrix determine the density of states
of the conductor
$N_{i} = \sum_{\gamma} \mbox{Tr} ({\cal N}^{(i)}_{\gamma\gamma})$;
the trace is over all quantum channels.
The non-diagonal elements are essential to describe fluctuations.

At equilibrium, if all contacts of conductor $i$ are held at the same
potential, the two conductors can be viewed, as the plates
of a capacitor holding a dipolar charge distribution
with an electrochemical capacitance\cite{btp}
$C_\mu^{-1} = C^{-1} + D_{1}^{-1} + D_{2}^{-1}$
which is the series combination of the
geometrical capacitance $C$ of the two conductors,
and the quantum capacitances $D_{i} = e^{2}{N}_{i}$
determined by their density of states. An excess charge
relaxes with a resistance determined by \cite{btp}
\begin{eqnarray}
    R^{(i)}_q &=& \frac{h}{2e^2} \frac{\sum_{\gamma\delta} \mbox{Tr}
    \left( {\cal{N}}^{(i)}_{\gamma\delta} {\cal{N}}^{(i) \dagger}_{\gamma\delta} \right)}
        {[\sum_{\gamma} \mbox{Tr}({\cal N}^{(i)}_{\gamma\gamma})]^{2}} .
        \label{rq}
\end{eqnarray}
In the presence of transport through the conductor $i$ the
role of the equilibrium charge relaxation resistance, Eq. (\ref{rq}),
is played by the non-equilibrium resistance\cite{plb} $R^{(i)}_v$,
\begin{eqnarray}
R^{(i)}_v  = \frac{h}{e^2} \frac{\mbox{Tr}
    \left( {\cal N}^{(i)}_{21} {\cal N}^{(i)\dagger}_{21}\right)}
    {[\sum_{\gamma} \mbox{Tr}({\cal N}^{(i)}_{\gamma\gamma})]^{2}}.
        \label{rv}
\end{eqnarray}
Note that the charge relaxation resistance $R^{(i)}_q$ invokes all
elements of the density of states matrix with equal
weight, but in the presence of transport the non-diagonal elements
are singled out. Next we relate
these resistances to the voltage fluctuations in the two coupled
mesoscopic conductors and subsequently to the dephasing time.
Here we mention only that at equilibrium, if all contacts
at each conductor are at the same potential, the dynamic
conductance of our capacitor is given by
$G(\omega) = - i \omega C_{\mu} + C_{\mu}^{2} R_q \omega^{2} + O(\omega^{3})$.
Thus $R_q$ determines the dissipation associated with
charge relaxation on the two conductors.

Charge and potential fluctuations are related by
\begin{equation}
\hat Q = C (\hat U_{1} - \hat U_{2}) =
e\hat{\cal N}_{1} - e^{2}N_{1}\hat U_{1}.
\end{equation}
\begin{equation}
- \hat Q = C (\hat U_{2} - \hat U_{1}) =
e\hat{\cal N}_{2} - e^{2}N_{2}\hat U_{2}.
\end{equation}
$\hat Q$ is the charge operator of the dipole.
These equations state that the dipole charge $Q$ on
conductors $1$ and $2$ can be written in two ways:
First it can be given by the potential differences and the geometrical
capacitances and second it can be expressed as sum of the bare
charges $e{\cal N}_{1}, e{\cal N}_{2}$ calculated in the absence of
screening and a screening charge which here is taken to be proportional
to the density of states of the conductor $N_{i}$ times the induced potential
$U_{i}$. Using $D_{i} = e^{2}N_{i}$ for the density of states
we find that the
effective interaction $G_{ij}$ between the two
systems is
\begin{equation}
    {\bf G} = \frac{C_{\mu}}{D_{1}D_{2}C} \left( \begin{array}{ll}

        (C+D_{2}) & C \\

        C & (C+D_{1})

        \end{array} \right)
        \label{geff} .
\end{equation}
With Eq. (\ref{geff}) we find for the potential operators
\begin{equation}
\hat U_{i} = e \sum_j G_{ij} \hat{ \cal N}_{j} .
\label{uop}
\end{equation}
The fluctuation spectra of the voltages
$S_{U_{i}U_{k}}(\omega) \delta(\omega + \omega^{\prime}) = 1/2
\langle {\hat U}_{i}(\omega){\hat U}_{k}(\omega^\prime) +
{\hat U}_{k}(\omega^\prime){\hat U}_{i}(\omega) \rangle$
now follow from the fluctuation spectra of the bare charges\cite{btp,plb},
\begin{eqnarray}
    S_{N_{i}N_{k}}(\omega) &=& \delta_{ik}
        \sum_{\delta\gamma} \int dE\, F_{\gamma\delta}(E,\omega) \nonumber\\
        && \mbox{Tr} [{\cal{N}}^{(i)}_{\gamma\delta}(E,E+\hbar\omega)
        {\cal{N}}^{\dagger (i)}_{\gamma\delta} (E,E+\hbar\omega) ].
\label{qfluct}
\end{eqnarray}
where $F_{\gamma\delta} = f_{\gamma}(E)(1-f_{\delta}(E+\hbar \omega))
+ f_{\delta}(E+\hbar \omega)(1- f_{\gamma}(E))$
is a combination of Fermi functions.
In the low frequency limit of interest here the elements of the
density of states matrix ${\cal{N}}^{(i)}_{\gamma\delta}$
are specified by Eq. (\ref{d0}). Using Eqs. (\ref{uop},\ref{su1})
we find that at equilibrium the low frequency fluctuations of the potential
in conductor $1$ are given by
\begin{equation}
S_{U_{1}U_{1}} = 2
(\frac{C_{\mu}}{C})^{2}
\left((\frac{C+D_{2}}{D_{2}})^{2} R^{(1)}_{q}
+ (\frac{C}{D_{1}})^{2} R^{(2)}_{q} \right) kT
\label{su1}
\end{equation}
with $R^{(i)}_{q}$ determined by Eq. (\ref{rq}).
Similar results hold for $S_{U_{2}U_{2}}$ and  the correlation
spectrum $S_{U_{1}U_{2}}$. If a bias $eV$ is applied, for instance to the
conductor $2$, we find the same spectrum as above, except that
$R^{(2)}_{q} kT$ is replaced, to first order in $e|V|$, 
by  $R^{(2)}_{v} e|V|$ for $e|V| > kT$.

To relate the voltage fluctuation spectra to the dephasing rate
we follow Levinson \cite{levinson}. A carrier in conductor $1$
moves in the fluctuating potential $U_{1}$.  As a consequence
the phase of the carrier is not sharp but on the average determined by
$\langle \exp (i(\hat{\phi}(t) - \hat{\phi}(0)) \rangle =
\langle \hat{T} \exp (i\int_{0}^{t} dt^{\prime}
\hat U_{1}(t^{\prime})) \rangle$. Assuming that the fluctuations are Gaussian
this quantum mechanical average is given by $\exp (-t/\tau_{\phi})$
with $\tau^{-1}_{\phi} = (e^{2}/2\hbar^{2}) S_{U_{1}U_{1}}$.
Since the voltage fluctuation spectrum Eq. (\ref{su1}) consists
of two additive terms we can decompose the dephasing rate into
two contributions $(1/\tau_{\phi})_{(11)}$ and $(1/\tau_{\phi})_{(12)}$
where the index pair $(ik)$ indicates that we deal with the
dephasing rate in conductor $i$ generated by the presence of conductor
$k$.

Before discussing the results it is useful to clarify the limit in which
we are interested. Typically, the Coulomb charging energy $U = e^{2}/2C$
is large compared to the level spacing $\Delta$ in the conductors
of interest. Since $\Delta_{i} = 1/N_{i}$ this has the consequence that
any deviations of the electrochemical capacitance from its geometrical
value are very small. We can thus take $C_{\mu} = C$
and ${C+D_{2}}/{D_{2}} \approx 1$ in Eq. (\ref{su1}).
Now we are interested in the
dephasing time $\tau^{(12)}_{\phi}$ in conductor $1$ due to the presence
of conductor $2$. Our discussion gives for this contribution
\begin{equation}
\left(\frac{1}{\tau_{\phi}}\right)_{(12)} = \frac{e^{2}}{\hbar^{2}}
\left(\frac{C}{D_{1}}\right)^{2} R^{(2)}_q kT .
\label{eqdep}
\end{equation}
with $R_q$ given by Eq. (\ref{rq}) if conductor $2$ is at equilibrium and
\begin{equation}
\left(\frac{1}{\tau_{\phi}}\right)_{(12)} = \frac{e^{2}}{\hbar^{2}}
\left(\frac{C}{D_{1}}\right)^{2} R^{(2)}_v e|V|,
\label{nondep}
\end{equation}
with $R_v$ given by Eq. (\ref{rv})
if it is in a transport state with $e|V| > kT$. 
Note that for 
closed $2D$-conductors $e-e$-scattering leads to a rate\cite{blanter} 
proportional to $T^{2}$
wheras for open conductors Eq. (\ref{eqdep}) predicts a rate 
which is linear in $T$. 

We now specify that transport in the conductor $1$
is via a single resonant tunneling state. Thus the relevant density
of states in conductor $1$ is a Breit-Wigner expression.
For simplicity we assume
that we are at resonance and hence $N_{1} = (2/\pi\Gamma)$, where
$\Gamma$ is the half-width of the resonance. The conductor $2$ is a
QPC. The resistance $R_q$ and $R_v$ for a QPC
in the absence of a magnetic field have been discussed in
Ref. \cite{plb}. $R^{(2)}_q$ is,
\begin{eqnarray}
    R^{(2)}_q &=& \frac{h}{e^2} \frac{\sum_{n} (d\phi_{n}/dE)^{2}}
    {[\sum_{n} (d\phi_{n}/dE)]^{2}}
\end{eqnarray}
where $\phi_{n}$ is the phase accumulated by carriers in the n-th 
eigen channel 
of the QPC traversing the region in which the potential
is not screened. Note that if only a single channel is open,
$R^{(2)}_q$ is universal
and given by $R^{(2)}_q = \frac{h}{e^2}$. Thus in the one-channel limit
the dephasing caused in conductor $1$ due to a QPC at equilibrium
is given by $(1/{\tau_{\phi}})_{(12)} =
({{\pi^{4}}{\Gamma^{2}}})/({hU^{2}}) kT$.

\begin{figure}
\narrowtext
\vspace*{0.5cm}
\epsfxsize=7cm
\centerline{\epsffile{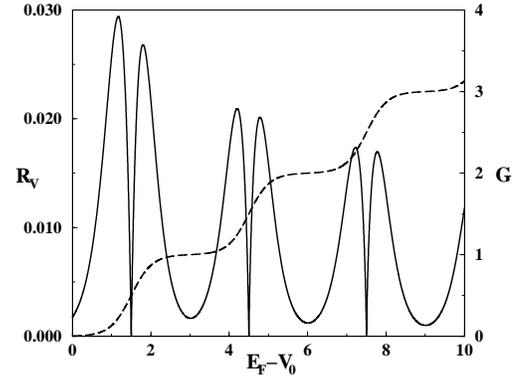}}
\vspace*{0.5cm}
\caption{ \label{rvedge}
$R_{v}$ (solid line) in units of ${h/e^{2}}$ and $G$ (dashed line)
in units of ${e^{2}/h}$ as a function of $E_{F} - V_{0}$ for a
saddle QPC with $\omega_x/\omega_y =1$ and $\omega_c/\omega_x = 4$.
}
\end{figure}

Next consider the case where a current is driven through
the QPC. The non-equilibrium charge relaxation resistance of
a QPC with transmission ${\cal T}_{n}$ and reflection 
probabilities ${\cal R}_{n}$ in the eigen channels is 
\begin{eqnarray}
R^{(2)}_v &=& \frac{h}{e^2} \frac{ \sum_n \frac{1}{{\cal T}_{n}{\cal R}_{n}}
        \left( \frac{d{\cal T}_{n}}{dE} \right)^2}{[\sum_{n} (d\phi_{n}/dE)]^{2}}.
        \label{rvqpc}
\end{eqnarray}
This result depends on the detailed shape of the QPC even in the single
channel limit. The similarity of this result with the one-channel (n =1)
result of Buks et al \cite{buks} can be seen by identifying the effective
variation $\Delta {\cal T}$ of the transmission coefficient with
$\Delta {\cal T} = (d{\cal T}_{1}/dE)(d\phi_{1}/dE)^{-1}$. Our result provides
a complete specification of the dephasing rate
in terms of the scattering matrix and geometrical capacitances.
We take screening into account and thus can
clarify the dependence on the quantum dot properties
(via $\Gamma$) and the capacitive coupling constant $C$.
$R_v$ has been evaluated for zero magnetic field in Ref. \cite{plb}.
For the high magnetic field case,
the non-equilibrium charge relaxation resistance for a QPC
is shown in Fig. \ref{rvedge}. We have used
a saddle\cite{bu90} point potential $V(x,y) = V_{0} - (1/2)m\omega_{x}^{2}x^{2}
+ (1/2) m \omega_{y}^{2}y^{2}$ and as in Ref. \cite{plb}
have evaluated the density of states semiclassically.
Note the strong suppression of the dephasing rate at threshold
of the opening of a new channel. Indeed the experiment of
Buks et al. \cite{buks} shows a double peak structure in the visibility
of the Ahronov-Bohm oscillations.

So far we always assumed that the QPC and
the dot are located
such that the total charge piled up in the QPC
matters. Thus the above results
involve only the energy derivatives of the
scattering matrix of the QPC and the dot.
Consider now the situation B shown in Fig. 1, where the quantum dot
is located away from the QPC down-stream along an edge.
Clearly, now the predominant interaction effect is due to the charge
fluctuations on the edge state adjacent to the quantum dot.
The charge which counts is that in a
region $\Omega_{B}$ affected by the potential
of the dot.

Very importantly, the approach introduced above can now be extended
to this more general situation. To generalize the above results
we need to find the charge and its fluctuations in region $\Omega_B$.
This can be accomplished by taking the derivative of the
scattering matrix with respect to a small
potential perturbation which extends over the region of interest.
Thus in general we arrive at a density of states matrix by replacing
the energy derivative in Eq. (\ref{d0}) by a functional derivative,
\begin{eqnarray}
d/dE \rightarrow  - \frac{1}{V_{\Omega}}\int_{V_{\Omega}} d^{3}{\bf r}
\frac{\partial}{\partial eU({\bf r})}
\end{eqnarray}
Let us apply this prescription to the case B of Fig. 1.

First, let us establish the scattering matrix for this system.
For the QPC with transmission probability ${\cal T} = 1- {\cal R}$
we chose 
$r \equiv s_{11} = s_{22} = - i {\cal R}^{(1/2)}$
and $t \equiv s_{21} = s_{12} = {\cal T}^{(1/2)}$. 
A carrier traversing the region $\Omega_{B}$ adjacent to the quantum dot
acquires a phase $\phi_2(U_2)$ where $U_2$ characterizes the potential
of the edge state in $\Omega_{B}$.  We assume that only the charge pile up
in the region $\Omega_{B}$ matters and consequently all additional phases
in the scattering problem are here without relevance. The total scattering
matrix of the QPC and the traversal of region $\Omega_B$ is then simply
$s_{11} = r, s_{21} = t, s_{12} = t \exp(i\phi_{2})$ and
$s_{22} = r \exp(i\phi_{2})$.

Consider next the charge operator. We have to evaluate the variation of the
scattering matrix with respect to the potential $U_{2}$.
Only $s_{12}$ and $s_{22}$ depend on this potential.
We find $ds_{12}/edU_2 = (ds_{12}/d\phi_2) (d\phi_2/edU_2)$.
But $(d\phi_2/edU_2) = - 2\pi dN_{2}/dE$, where now $dN_{2}/dE$ is
the density of
states of the edge state in region $\Omega_{B}$ of conductor $2$.
Simple algebra
now gives ${\cal N}^{(2)}_{11} =  {\cal T} dN_{2}/dE$,
\begin{equation}
{\cal N}^{(2)}_{12} = {\cal N}^{(2) \ast}_{21} = r^{\ast}t
\exp (-i\phi_2) dN_{2}/dE,
\label{n12}
\end{equation}
and ${\cal N}^{(2)}_{22} = {\cal R} dN_{2}/dE$.
At equilibrium we find $R^{(2)}_q = h/2e^{2}$ as
is typical for an edge state
that is perfectly connected to a reservoir\cite{christen}.
The non-equilibrium resistance is
\begin{equation}
R^{(2)}_{v}= (h/e^{2}) {\cal T}{\cal R}.
\label{rvqhe}
\end{equation}
Note that in the one-channel
case both $R_q$ and $R_v$ are independent of the density of states
$N_{2}$. The additional dephasing rate generated by the edge at
equilibrium in the quantum dot at resonance is
$(1/{\tau_{\phi}})_{(12)} = ({{\pi^{4}}{\Gamma^{2}}}/{2hU^{2}}) kT$.
Note that this rate
depends on the edge state only through its geometrical capacitance.
In the non-equilibrium case, the additional dephasing rate caused
by the charge fluctuations on
the edge state is
$(1/{\tau_{\phi}})_{(12)} = ({{\pi^{4}}{\Gamma^{2}}}/{hU^{2}}) 
{\cal T}{\cal R}e|V| .$
A rate proportional to ${\cal T}{\cal R}$ is also obtained by 
Buks et al. \cite{note1}. Of interest is the effect of screening: 
While in the one channel case, the rate depends on the capacitance of the 
edge channel only, such a universal result does not apply as soon as 
additional edge states are present. Thus consider an additional
edge state which is transmitted with probability $1$.
It generates no additional noise and leaves the dc-shot
noise invariant\cite{bu90}.
But the additional edge channel contributes to screening.
If we take the two edge channels to be close together in the region
$\Omega_{B}$ both edge channels will see the same potential $U_{2}$.
Now the total density of states of the two edge channels in
region $\Omega_{B}$ has a contribution from both the perfectly
transmitted edge state (1) and edge state (2)
$N_{2} = N_{21}+ N_{22}$.
As a consequence the dephasing rate is now reduced and given by
\begin{equation}
\left(\frac{1}{\tau_{\phi}}\right)_{(12)} = \frac{{\pi^{4}}{\Gamma^{2}}}{hU^{2}}
\left(\frac{N_{22}}{N_{21}+ N_{22}}\right)^{2} {\cal T}{\cal R}e|V| .
\label{last}
\end{equation}
Eq. (\ref{last}) is valid if there is no population equilibration
among the two edge channels between the QPC and the dot.
If there is equilibration (which can be achieved by placing
a voltage probe between the QPC and the dot\cite{note1}) 
we will show elsewhere
that the dephasing rate becomes
$({1 / \tau_{\phi}})_{(12)} = ({\pi^{4}}{\Gamma^{2}}/{hU^{2}\nu^{2}})
{\cal T}{\cal R}e|V|$ where $\nu$ is the number of edge states.
In particular if only one edge state is present the dephasing rate
is unaffected by the presence of a phase randomizing reservoir.
This result provides a simple test of the theory presented here.

If the magnetic polarity is reversed the additional dephasing rate
is only due the equilibrium fluctuations, independent of whether or not
a bias is applied to the QPC.

In this work we have presented a discussion of the dephasing
in Coulomb coupled mesoscopic conductors which is based on
the fluctuations of the dipolar charge that is generated
by the long range Coulomb interaction. This dipole
is associated with a capacitance
and its dissipative behavior is characterized by charge relaxation
resistances $R_q$ and $R_v$.
These resistances are determined by the low frequency collective
modes of the Coulomb coupled conductors.

This work was supported by the Swiss National Science Foundation
and by the TMR network Dynamics of Nanostructures.\\
\vspace{-.5cm}

\end{multicols}

\begin{references}

\vspace{-1cm}

\bibitem{field}  M. Field, et al. Phys. Rev. Lett. {\bf 70}, 1311 (1993).

\bibitem{buks}   E. Buks, et al. Nature {\bf 391}, 871 (1998).

\bibitem{aleiner1}  I. L. Aleiner, N. S. Wingreen, and Y. Meir,
                    Phys. Rev. Lett. {\bf 79}, 3740 (1997).

\bibitem{levinson}  Y. Levinson, Europhys. Lett. {\bf 39}, 299 (1997).

\bibitem{aak}       B. L. Altshuler, A. G. Aronov and D. Khmelnitskii,
                    J. Phys. C{\bf 15}, 7367 (1982).

\bibitem{blanter}   U. Sivan, I. Imry and A. G. Aronov, Europhys. 
                    Lett. {\bf 28}, 115 (1994); Y. M. Blanter, 
                    Phys. Rev. B{\bf 54}, 12807 (1996).

\bibitem{btp}       M. B\"{u}ttiker, H. Thomas, and A. Pretre,
                    Phys. Lett. {\bf A180}, 364, (1993).

\bibitem{bu90}      M. B\"{u}ttiker, Phys. Rev. Lett. {\bf 65}, 2901 (1990);
                    Phys. Rev. B{\bf 41}, 7906 (1990).

\bibitem{plb}       M. H. Pedersen, S. A. van Langen and M. B\"{u}ttiker,
                    Phys. Rev. {\bf B 57}, 1838 (1998).



\bibitem{mohant}  P. Mohanty, E. M. Q. Jariwalla and R. A. Webb, 
                  Phys. Rev. Lett. {\bf 79}, 3306 (1997). 
    

\bibitem{zawad}   A. Zawadowski, Jan von Delft, D. C. Ralph, 
                  cond-mat/9902176
              
\bibitem{huibers} A.G. Huibers, et al. Phys. Rev. Lett. 81, 200 (1998).

\bibitem{brouwer} P. W. Brouwer and M. B\"uttiker, 
                  Europhys. Lett. {\bf 37}, 441-446 (1997).
                            
\bibitem{note1}   E. Buks, D. Sprinzak, M. Heiblum, D. Mahalu, V. Umansky and 
                  H. Shtrikman, (unpublished). 
 
\bibitem{smith}   F. T. Smith, Phys.\ Rev.\ {\bf 118} 349 (1960).
                  
\bibitem{note2}   Y.~V.\ Fyodorov and 
                  H.~J.\ Sommers, Phys. Rev. Lett. {\bf 76}, 4709 (1996);
                  V.~A.\ Gopar, P.~A.\ Mello, and M.\
                  B\"{u}ttiker, Phys.\ Rev.\ Lett.\ {\bf 77}, 3005 (1996);  
                  P.~W.\ Brouwer, K.~M.\ Frahm, and C.~W.~J.\ Beenakker, 
                  Phys.\ Rev.\ Lett.\ {\bf 78}, 4737 (1997);
                  C. Texier and A. Comtet, cond-mat/981219. 
                               
\bibitem{christen}  M. B\"{u}ttiker and T. Christen, in 
                   'High Magnetic Fields in the Physics of Semiconductors', 
                    edited by G. Landwehr and W. Ossau, (World Scientific, 
                    Singapur, 1997). p. 193. cond-mat/9607051
                               
\end{references}
\end{document}